# Definition and evaluation of a finite element model of the human heel for diabetic foot ulcer prevention under shearing loads

**Authors :**

Alessio Trebbi*
Univ. Grenoble Alpes, CNRS, UMR 5525, VetAgro Sup, Grenoble INP, TIMC, 38000 Grenoble, France.
Yohan.Payan@univ-grenoble-alpes.fr

Nolwenn Fougeron
Univ. Grenoble Alpes, CNRS, UMR 5525, VetAgro Sup, Grenoble INP, TIMC, 38000 Grenoble, France.
nolwenn.fougeron@hotmail.fr

Yohan Payan
Univ. Grenoble Alpes, CNRS, UMR 5525, VetAgro Sup, Grenoble INP, TIMC, 38000 Grenoble, France
Yohan.Payan@univ-grenoble-alpes.fr

*Corresponding author

**Abstract**

Diabetic foot ulcers are triggered by mechanical loadings applied to the surface of the plantar skin. Strain is considered to play a crucial role in relation to ulcer etiology and can be assessed by Finite Element (FE) modelling. A difficulty in the generation of these models is the choice of the soft tissue material properties. In the literature, many studies attempt to model the behavior of the heel soft tissues by implementing constitutive laws that can differ significantly in terms of mechanical response. Moreover, current FE models lack of proper evaluation techniques that could estimate their ability to simulate realistic strains.

In this article, we propose and evaluate a FE model of the human heel for diabetic foot ulcer prevention. Soft tissue constitutive laws are defined through the fitting of experimental stretch-stress curves published in the literature. The model is then evaluated through Digital Volume Correlation (DVC) based on non-rigid 3D Magnetic Resonance Image Registration. The results from FE analysis and DVC show similar strain locations in the fat pad and strain intensities according to the type of applied loads. For additional comparisons, different sets of constitutive models published in the literature are applied into the proposed FE mesh and simulated with the same boundary conditions. In this case, the results in terms of strains show great diversity in locations and intensities, suggesting that more research should be developed to gain insight into the mechanical properties of these tissues.

## 1. Introduction

One out of four adults with diabetes develops a diabetic foot ulcer in his/her lifetime [1][2]. The cost to medicate chronic wounds is estimated in US at $ 28.1–96.8 billion per year [3]. Diabetic foot ulcers can be triggered by sustained mechanical loading, normal pressure and/or shear forces on the skin and underlying soft tissues, especially over bony prominences [4]. The incidences of diabetic foot ulcers are still too high considering that two-fifth of patients being taken in charge in a geriatric unit will develop a pressure ulcer [5][6][7]. Therefore, the development of innovative strategies to prevent diabetic foot ulcer is required. To successfully prevent such an ulcer, healthcare professionals should be able to predict if a certain state of mechanical loading (such as a load applied on the skin surface) would lead to internal localized irreversible tissue necrosis [8][9].

Finite Element (FE) modelling is a technique that is considered to have the potential to assist the prevention of diabetic foot ulcers either as a tool for preclinical testing of support interfaces and/or identifying anatomical risk factors [10][11]. This solution is a conventional computational technique for estimating internal stress and strains derived by the application of a surface loading. This information can then be used to predict tissue damage where these exceed appropriate thresholds [12].

Soft tissues can be loaded perpendicularly, described clinically as pressure, or in a direction parallel to the skin surface, described as shear. Nowadays it is considered that a combination of these two loadings could result (I) in occlusion of blood and lymphatic vessels which is suspected to be particularly damaging [13] and (II) the weakening of the cellular membrane of the soft tissues cells [14]. It is considered that shear strains could play a crucial role in the development of deep tissue injuries [15][16]. It is therefore believed that a FE model should properly simulate the behavior of soft tissues under shearing loads, in order to successfully predict the formation of diabetic foot ulcers. A common methodology to simulate the heel in a loaded configuration is to apply an imposed displacement on the skin surface and analyze the resulting strains [17][18]. Injury criteria include maximum values of strains, averaged over a volume (called "cluster" [19]) in which a specific threshold is exceeded.

Experimental data on injury thresholds was reported by Ceelen et al. in an animal study which demonstrated that the Green-Lagrange maximal shear strain was the best predictor of pressure injury [20]. In a subsequent study, the team demonstrated that these strains could be estimated with FE modelling [21]. However, the design of personalized FE model is still an open challenge, more particularly considering the characterization of the constitutive behavior of each soft tissue included in the model, such as skin, fat, muscles [22]. As the boundary conditions implemented in the models published in the literature are different, the results provided by such models are difficult to compare [23][24]. When considering studies involving FE models of the heel, researchers adopt stiffness parameters from curve fitting data from cadavers or animals experiments [25][26], inverse modelling [27], or values extracted from other published papers [28]. In the literature, data on soft tissue mechanical properties is diverse. For the same material, it is possible to find studies reporting stiffness parameters that differ of several orders of magnitude [28][29]. This can be misleading and confusing when researchers have to select a specific set of constitutive laws and parameters for a FE model to simulate a specific configuration. At this point, it is clear that new tools are required to evaluate these FE models.

FE analyses require a consistent and robust methodology for evaluation, involving rigorous experimental measurements [30]. In the literature, most studies establish the validity of FE models via experimental measurements. Concerning foot modeling, researchers proposed to evaluate FE models in terms of their capacities to simulate interface plantar pressure [31]. However, Macron et al. has shown that interface pressure distributions do not link directly with internal strains [11]. Therefore, pressures only cannot be used to predict the internal strains. Linder-Granz et al. compared the soft tissue contours geometry between the FE simulation and the segmentation results [32]. This solution, however, considers only the external geometry and not the measurement of interest, which are the local internal tissue strains. Probably the most crucial aspect is that, to the best of our knowledge, none of these models were evaluated with any technique to analyze specifically their performances under shearing load conditions.

In a recent work, Trebbi and colleagues proposed a Digital Volume Correlation (DVC) technique based on volumetric Image Registration (IR) of Magnetic Resonance (MR) images to calculate internal strains from the application of shearing loads [33]. This solution is based on finding the non-rigid transformation that matches two MR images of the same body part, one in an unloaded configuration and the other one in a mechanically loaded configuration [34]. Tissue strains in the loaded deformed configuration are then computed from that non-rigid transformation field. This solution offers a methodology to quantitatively evaluate the strains that are simulated from a FE model deformed with the same loading configuration.

The objective of this paper is to define and evaluate a FE model of the human heel for diabetic foot ulcer prevention. To this purpose, a 3D subject-specific FE model of the human heel is elaborated. Regarding the evaluation, it is believed to be crucial to analyze the behavior of soft tissues under shearing loads configurations, as these are considered the most dangerous for triggering diabetic foot ulcers. The strains computed by the FE simulations under shearing loads configurations are compared with the strains measured by the IR methodology proposed by Trebbi et al. [33]. In parallel, another type of analysis is developed by comparing the simulation results with other constitutive laws proposed in the literature for the definition of heel FE models. This is done to analyze the constitutive behavior of the proposed model with respect to the other ones already published in the literature.

## 2. Methods

**Imaging**

The MR images used in this paper have been acquired in a previous work. The details regarding the experimental setup are summarized in this paragraph. For a deeper insight, the reader is referred to the related article [34]. A healthy volunteer (male, 40 years old) gave his informed consent to participate in the experimental part of a pilot study approved by an ethical committee (MammoBio MAP-VS pilot study N°ID RCB 2012-A00310-43, IRMaGe platform, Univ. Grenoble Alpes). The right foot of the subject was locked in the loading device designed to apply a static normal force and a shearing force on the heel pad by means of an indenting platform. Five loading configurations, including normal and/or shearing loads (Table 1), were defined in order to capture the non-linear mechanical properties of the soft tissues. The setup is illustrated in Figure 1. A double face tape was applied on the plate surface to allow the application of the shearing load without any slipping. A proton density MR sequence was used to obtain 3D images while the loadings were applied. The five 3D MR considered images of the heel are listed in Table 1.

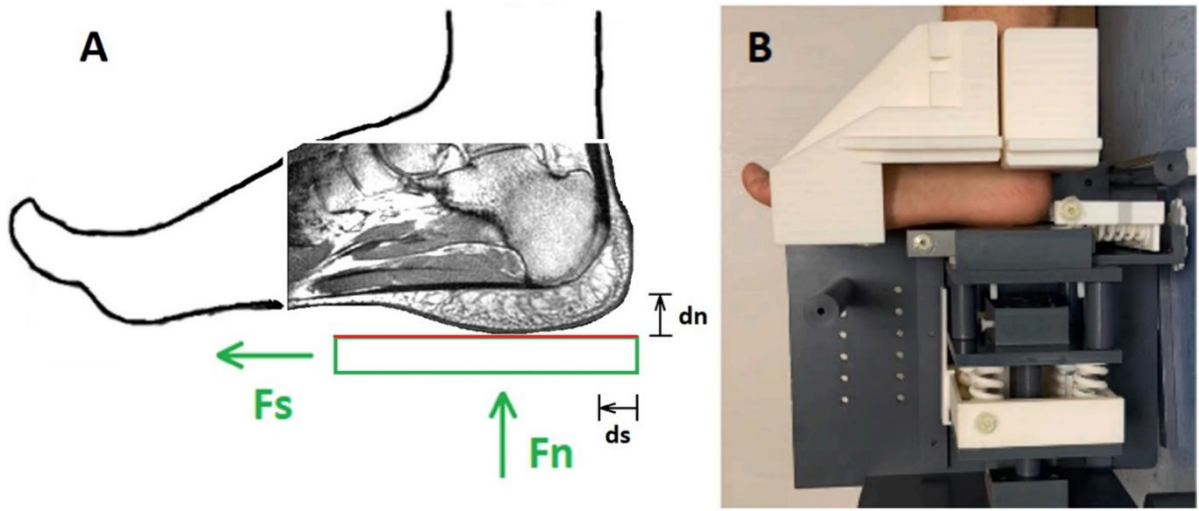

*Figure 1: A) Schematic representation of load application on the heel in the MR acquisition. The green rectangle represents the loading platform and the respective arrows show the application of forces (Fn normal force, Fs shearing force). The plate displacements (dn normal displacement, ds shear displacement) are shown in black. The double-face tape is shown in red. B) Internal view of the loading device* [34]*.*

| Name | Description | Load |
|---|---|---|
| Load0 | Unloaded heel | 0 N |
| Load1 | Heel with high normal load | 140 N |
| Load2 | Heel with normal load | 15 N normal |
| Load3 | Heel with normal and high shearing load | 15 N normal and 45 N shear |
| Load4 | Heel with normal and shearing load | 15 N normal and 12 N shear |

*Table 1 : List of lading configurations*

A 2D sagittal snapshot of the MR volume of Load0, Load1 and Load3 is provided in Figure 2: Sagittal MR slices for the loading configurations Load0-unloaded, Load1-normal load, Load3-shering load..

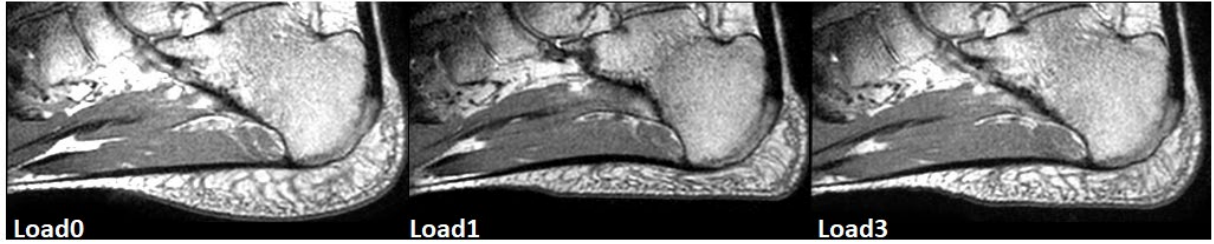
*Figure 2: Sagittal MR slices for the loading configurations Load0-unloaded, Load1-normal load, Load3-shering load.*

**Generation of the heel 3D FE mesh**

Manual segmentation of MR images was performed using Amira 6.5.0 (Thermo Fisher Scientific Inc.). The unloaded Load0 acquisition was used to segment the fat pad, tendon, muscles and skin. The resulting surface files were imported in HyperMesh 2019 (Altair Engineering, Inc.) for FE mesh generation. All bones of the foot and tissues above were not included in the model in order to reduce computation time. This 3D mesh (Figure 3) was composed by 80,000 linear tetrahedral elements (with a mesh size of 2 mm). The 3D mesh was then imported into ANSYS 19.2 APDL (ANSYS, Inc., Canonsburg, PA). A rectangular platform was simulated underneath the FE model of the heel representing the plate of the loading device. The region selected for the model was chosen to be large enough to include all the tissue displacements generated by the plate [33]. A mesh convergence study was performed to verify the consistency of the simulations for the adopted mesh density.

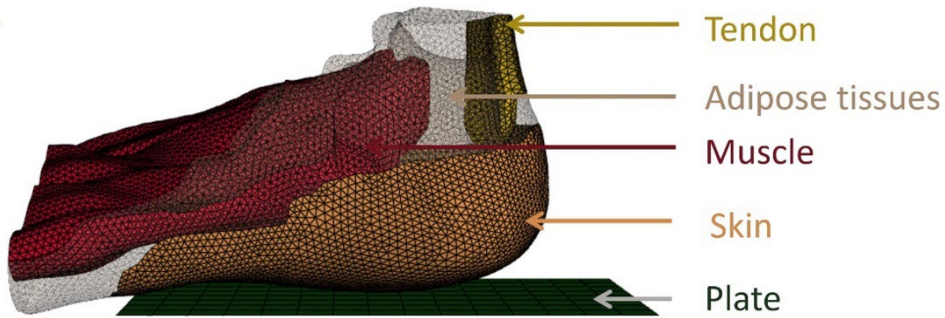
*Figure 3: Finite element mesh of the heel generated from MR image segmentation*

**Material properties**

The calcaneus bone was assumed to be rigid as, for the considered application, its deformations are neglectable in comparison with the ones of the surrounding soft tissues [35]. The boundary between the bones and soft tissues were modeled as a rigid surface. The skin was modelled with an Isihara's et al. law (equivalent to a Yeoh constitutive law with the parameter $C_{30}$ equals to zero) [36]. Material parameters were optimized using a curve fitting method with Matlab (MathWorks Inc, USA) from the skin experimental data of Annaidh et al. [37]. Muscles were modeled using a Yeoh constitutive law with the same procedure using the data of uniaxial tensile test from Gras et al. [38]. Considering the fat pad, a first order Ogden constitutive law was chosen as proposed by Moerman et al. [39]. Material parameters were optimized using the same procedure according to the quasi-static experimental data from Miller-Young et al. who performed several compression tests on the foot fat tissues [40]. Table 2 gathers all the corresponding constitutive parameters values for skin, fat and muscle. Incompressibility parameters in all directions were supposed equal assuming a Poisson ratio $\nu$ of 0.4999 [41]:

| Tissue | ν | µ (MPa) | α | $C_{10}$ (MPa) | $C_{20}$ (MPa) | $C_{30}$ (MPa) |
|---|---|---|---|---|---|---|
| Fat | 0.4999 | 0.0034 | 6.2 | | | |
| Skin | 0.4999 | | | 0.2650 | 1.9230 | - |
| Muscle | 0.4999 | | | 0.0050 | 0.0690 | 1.9670 |

Table 2 : Set of constitutive laws implemented in the FE model of the heel. With v, the Poisson ratio, set to 0.4999, to account for the nearly incompressibility of soft tissues.

In parallel, isotropic constitutive laws from the literature related to FE heel models were implemented in our heel FE mesh to compare the results of the corresponding simulations. Table 3 lists the six hyper-elastic material laws and their respective parameters that are compared here.

| Study | Tissue | Method | Model | v | α | µ (kPa) |
|---|---|---|---|---|---|---|
| Akrami [42] | Homogeneous | Average | LE | 0.49 | | 385.9 |
| Edemir [27] | Homogeneous | Inverse modelling | OG | 0.49 | 6.82 | 16.45 |
| Friedman [28] | Skin | Literature [43] | NH | 0.495 | | 324.7 |
| | Fat | Literature [44] | NH | 0.495 | | 0.290 |
| Luboz [45] | Skin | Average | NH | 0.495 | | 66.89 |
| | Fat | Average | NH | 0.49 | | 10.07 |
| | Muscle | Average | NH | 0.495 | | 20.07 |
| Spears [44] | Skin | Inverse modelling | OG | 0.49 | 6.80 | 640 |
| | Fat | Fitted | OG | 0.495 | 8.80 | 0.290 |
| Zwam [29] | Skin | Literature [45] | OG | 0.49 | 2.3 | 50 |
| | Fat | Literature [27] | OG | 0.49 | 6.82 | 16.45 |
| | Muscle | Literature [45] | LE | 0.49 | | 20.07 |

Table 3: List of constitutive models presented in the literature for the soft tissues of the human heel. LE = Linear elastic, OG = Ogden, NH =Neo-Hookean. The coloring on the tissue column are made in order to facilitate comparisons between the same tissues.

**Boundary conditions**

Boundary conditions were defined based on the MR load configurations. The upper boundary of the foot which is the limit between the bones, not included here, and the soft tissues was considered as rigid. All nodes' degrees of freedom were constrained by the degrees of freedom of the barycenter of these nodes, computed with equal weights for all nodes. The barycenter was fixed during the analysis so its degrees of freedom were all set to zero. The simulations were divided in two steps. First, the normal load was applied on the plate and second, the shearing load was added. Contacts between soft tissue components (skin, fat, muscle) were assumed without sliding. As soon as the collision was detected, the contact between the plate and the skin was also assumed as bounded with no slip since a tape was placed at this interface during the MR acquisitions.

**Image Registration**

The MR image registration technique presented in this article has been developed in a previous study [33]. The main details regarding the steps to perform the IR analysis are summarized in the following paragraph. For additional details concerning the trueness and repeatability of the measurements, the reader is referred to the associated publication. DVC was implemented using the publicly available Elastix package used to non-rigidly register 3D MR images [46]. The four MR loaded images (Load1-Load4) of the heel were first rigidly registered to align the calcaneus bone with the unloaded image (Load0). Then, the image-based non-rigid registration was used to estimate the displacements and Green Lagrange maximal shear strain fields [12].

## 3. Results

In order to evaluate each FE model (the one we propose and the six other models built with constitutive laws from the literature), the results from the simulations are compared with the outputs from the IR that are considered as a reference. First, the intensity of strains is discussed by analyzing the volume of tissue subjected to a specific amount of strain; this type of representation is indeed used in strain analysis for diabetic foot ulcer prevention[19][47][48]. Second, the location of strain concentrations is analyzed by inspecting sagittal sections that show the levels of strains in the internal layers.

*Strain intensity*

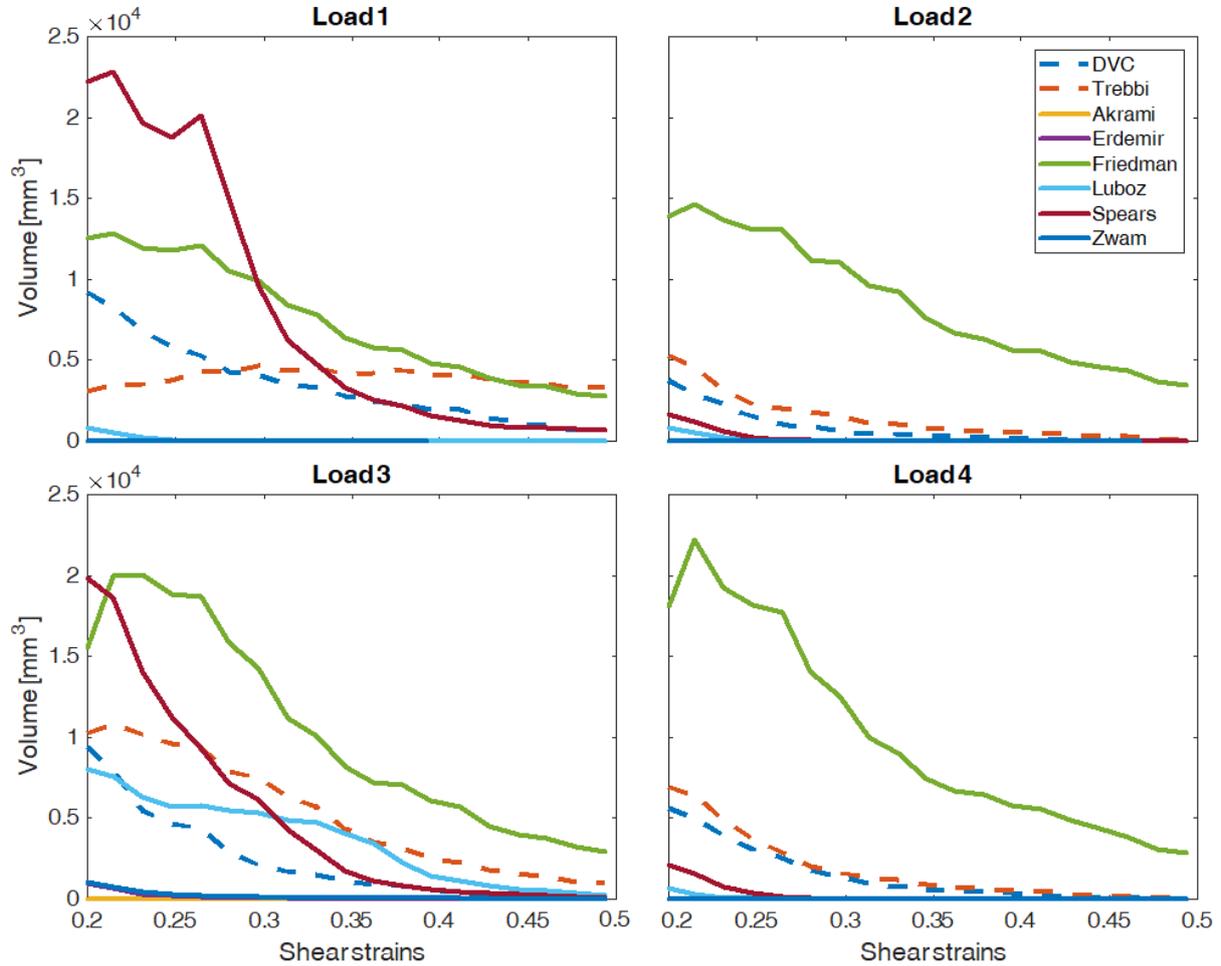

*Figure 4: Volume of soft tissue subject to a specific amount of Green Lagrange shear strain intensity.*

| Model/Load | Trebbi | Luboz | Eedemir | Zwam | Akrami | Spears | Friedman |
|---|---|---|---|---|---|---|---|
| Load1 | 0.42 | 0.65 | 0.59 | 0.66 | 0.66 | 0.91 | - |
| Load2 | 0.05 | 0.20 | 2.17 | 0.21 | 0.21 | 0.18 | 1.40 |
| Load3 | 0.55 | 0.27 | 0.41 | 0.40 | 0.44 | 0.20 | 1.52 |
| Load4 | 0.07 | 0.26 | 0.27 | 0.27 | 0.27 | 0.22 | 1.63 |
| Total | 1.11 | 1.39 | 1.50 | 1.57 | 1.60 | 1.51 | >4.55 |

*Table 4 : For each curve plotted on figure 4 and representing a given constitutive model, the area located between this curve and the reference DVC curve is computed. The higher the area is, the larger is the error. Constitutive models are ordered from the lowest total error to the highest one.*

In this section, the amount of tissue under a specific intensity of strain is considered. The focus is given for strain levels comprised between 0.2 and 0.5. This was chosen for two reasons. First, the intensity of strains falls in this range for most of the configurations. Second, these values are the base levels of strain considered to be possible cause of tissue damage [12]. As proposed by Grigatti & Gefen [48], the volume of tissue subject to a specific amount of strain is plotted on Figure 4 for the seven constitutive models studied here, as well as for the reference data measured with DVC. For each curve plotted on this figure and representing a given constitutive model, the area located between this curve and the reference DVC curve is computed. The higher the area is, the larger is the error. Table 4 summarizes the errors for each constitutive model and for each load configuration.

With this criteria, the constitutive law proposed in this paper (called *Trebbi* here) seems to be the best to describe the strain intensities across the considered loading configurations (Table 4). The best fits are related to the configurations with the lower amount of load, namely Load2 and Load4. This emphasizes the complexity of modeling hyperelastic tissues that vary significantly their stiffness in relation to the amount of strain they experience.

The model of Friedman showed high levels of strains distributed in a large amount of volume. This is due to the low value of shear modulus implemented for the fat 0.290 kPa (Table 3). The same value was used by Spears, but in this case, a significant higher stiffness for the skin was selected. This resulted in a significant reduction of the strains for the lower loading configurations Load2 and Load4. The impact of shearing loads can be seen in the results from the model of Luboz. For Load3 (with 45 N of shearing load) the model had a considerable higher amount of shearing strains compared with Load1 (140 N of normal load). The models of Zwam, Erdemir and Akrami showed a stiff behavior that maintained the strain propagation low compared to the DVC results.

*Strain location*

In this section, the location of strain concentrations is analyzed by looking at a sagittal slice passing through the lowest point of the calcaneus. This slice was selected as experiments have shown that the highest values of strains are mainly located in this area. Results are plotted in Figure 5.

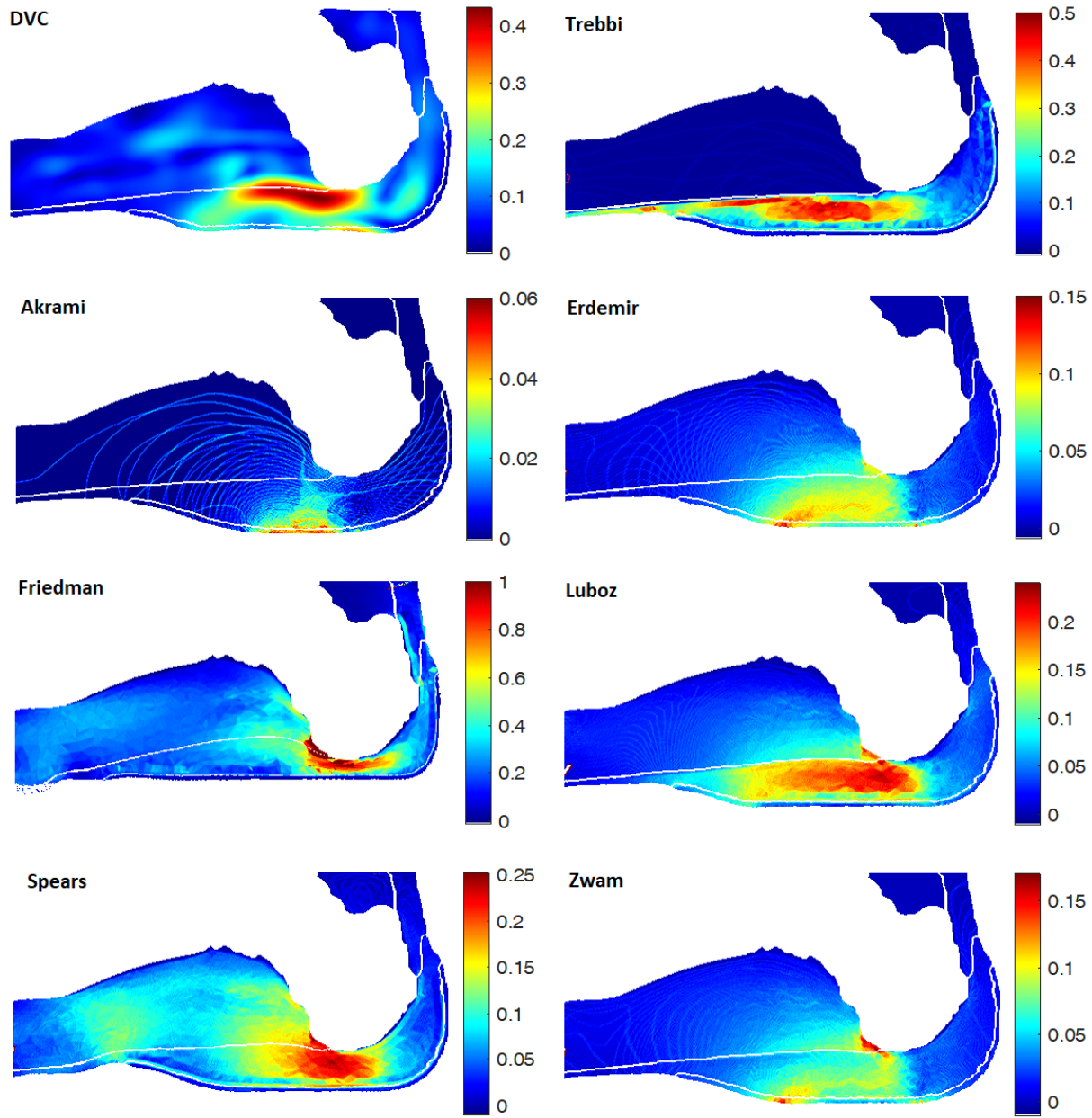

*Figure 5: Green Lagrange strains propagation for all the constitutive models in loading configurations Load4 (15 N normal and 12 N shear forces).*

The comparison shows that Luboz, Spears and Trebbi models provide strain concentrations in comparable locations with the DVC, namely in the fat pad around the calcaneus head with a propagation towards the forefoot under the aponeurosis. Considered the proposed model (Trebbi) the strains keep propagating in the interface between the muscle and the fat. This is due to the different compliance between the constitutive laws of these two tissues. With this respect, the muscle appears probably to be too stiff here considering the uniform low levels of strains in the muscular region.

The selection of a stiffer behavior for the skin can be clearly seen in the models of Spears and Trebbi as there is a sudden change in the strain propagation in the interface between the fat and the adipose tissue. From the deformed shapes of the image, it is possible to have an idea of the amount of plate displacement resulted from the simulation, with Trebbi and Spears models that seem the closest to the current plate displacement. The overall stiffness of the tissues has relevant effects in the strain locations as in the cases of Akrami and Friedman. For Akarmi (too stiff) strains are concentrated on the contact region between the plate and the skin. The amount of displacement applied by the plate is low with the skin being almost in its undeformed configuration. On the other hand, in Friedman's model (too compliant), the strains are unrealistically high and strongly concentrated around the head of the calcaneus. In that case, the fat pad is completely flattened by the load applied by the plate.

It must be noted that the intensity of strain shown in the considered slice can be misleading as this representation is showing the values of this slice only (Figure 5). Strain values can vary significantly between the slices. Therefore, in relation to the magnitude of strains, the reader is addressed to the strain intensity paragraph.

## 4. Discussion

The objective of this paper was to define and evaluate a FE model of the heel for foot ulcer prevention under shearing loads. To this account, a FE mesh was generated from a MR image and a set of constitutive equations was derived from curve fitting experimental results from the literature. To evaluate the model, the results from the simulations were compared with data obtained from a MR experiment applying different intensities of normal and shearing loads. IR was used in combination with the MR images of loaded tissues to estimate the generated internal displacements and strains. This technique allowed to gain insights on the location of the shear strain clusters and their intensity. It allowed also a quantitative comparison with the strains computed by the FE simulation. The FE results obtained with the selected constitutive equations were compared with other constitutive laws reported in the literature for other FE heel models. Six sets of material models from the literature were therefore considered to run simulations with the same boundary conditions as for our FE model. The differences in terms of strain intensity and propagation were analyzed.

Results showed larger discrepancies between data and FE simulations for high loads (Load1, Load3). This is due to the hyperelastic behavior and heterogeneous/anisotropic structure of the fat pad, which stiffness varies significantly in relation to the amount of load applied. The constitutive laws proposed in this article provided the best results as compared to the strains measured by DVC. This is particularly encouraging since we know that some improvements should be done to better fit the current subject-specific constitutive parameter, which would probably require the use of in vivo measurements such as elastography [49] or local aspiration [50].

As concerns the comparison with other literature constitutive models, results show that the choice of the material properties is critical for the analysis as it may provide highly different strains patterns among models. The constitutive parameters have also a significant impact on the location and the intensity of the strain concentrations. This confirms the fact that it is crucial to implement constitutive laws that refer specifically to the considered subject when building a subject-specific FE model. MR compatible experiments such as the ones presented in Gefen et al. [51], Spears et al. [44] or the one introduced in this paper, offer space to estimate in vivo material properties [52][53].

It is important to note that the methodology introduced in this paper, to compare strains computed by FE models and strains measured by DVC, is quite new. It is therefore difficult to compare such a methodology with other methods proposed in the literature. However, some considerations can be made on strain intensities and locations found in previous studies. Zwam et al. looked at strains generated into the deep tissues of the heel for a subject laying in a supine position. The analyzed Green Lagrange strains reached a level of 0.4 that is comparable with what was observed in the FE simulations [29]. As concerns strain locations, our results are in line with what was found in previous studies, namely high strains located in the fatty region, close to the bony prominence [28], [54].

The current work still suffers from some limitations. First, the DVC and all the FE models were defined from the MR images of a single subject. This limits the implemented geometry of the FE models to a single case without account for the impact of morphological variability. Second, the considered loadings are static and therefore ignore the viscoelastic properties of the tissues. This is an important property of biological tissues and it should be considered when attempting to specifically prevent diabetic foot ulcers. The future steps will aim to take into account this property in the simulations in combination with DVC measurements. Third, the implemented constitutive laws were obtained from ex vivo experiments proposed in the literature. This is clearly a limitation since it is known that subject-specific constitutive parameters differ from the one measured on cadavers. The in vivo estimation of such parameters should therefore be explored in future studies.

## 5. Conclusion

This paper introduced an original methodology to compare tissue strains computed by various FE constitutive models of the human heel with strains estimated from MR-based DVC. This technique allowed for the definition of a FE model of the human heel to be implemented in diabetic foot ulcer preventions. To this account, the model can be used to simulate foot/shoe interactions to analyze the impacts in terms of strains resulting from the interface pressures. These simulations could allow to optimize insole design to reduce the risk of diabetic foot pressure ulcers. The methodology could easily be implemented in other body locations such as sacrum or buttocks where strain propagation is fundamental to understand pressure ulcer risks. The results provided in this paper should also be considered as an illustration of the importance for choosing the constitutive equation for each tissue implemented in the FE models, especially for larger strains when hyperelastic laws are considered. This is of crucial importance as material parameters proposed in the literature can range over several orders of magnitude for the same tissue. Besides the evaluation provided by our methodology, the final validation of the ability of FE models to predict pressure ulcers will only be given through clinical trials aiming to quantify tissue damage and ulcer prediction. Until this last step, the effectiveness of biomechanical modeling for diabetic foot ulcer prevention remains limited.

## Conflict of interest
None

## Acknowledgments

This research has received funding from the European Union's Horizon 2020 research and innovation programme under the Marie Skłodowska-Curie Grant Agreement No. 811965; project STINTS (*Skin Tissue Integrity under Shear*).